\documentclass[aps,prb,reprint,groupedaddress,amsmath,amssymb,superscriptaddress]{revtex4-1}

\usepackage{graphicx}
\usepackage{dcolumn}
\usepackage{amsmath,amsthm}
\usepackage{epstopdf}
\usepackage{color}
\usepackage{ulem}
\usepackage{ gensymb }

\begin{document}
	
	
	\title{Spin control of light with hyperbolic metasurfaces}
	
	
	
	\author{Oleh~Y. Yermakov}
	\email{o.yermakov@phoi.ifmo.ru}
	\author{Anton~I. Ovcharenko}	
	\author{Andrey~A. Bogdanov}
	\author{Ivan~V. Iorsh}	
	\affiliation{ITMO University, St.~Petersburg 197101, Russia}	
	\author{Konstantin~Y. Bliokh}
	\affiliation{Center for Emergent Matter Science, RIKEN, Wako-shi, Saitama 351-0198, Japan}	
\affiliation{Nonlinear Physics Centre, Research School of Physics and Engineering,
Australian National University, Canberra ACT 2601, Australia}
	\author{Yuri~S. Kivshar}
	\affiliation{ITMO University, St.~Petersburg 197101, Russia}	
	\affiliation{Nonlinear Physics Centre, Research School of Physics and Engineering,
Australian National University, Canberra ACT 2601, Australia}

	
	
	\date{\today}
	
	\begin{abstract}
Transverse spin angular momentum is an inherent feature of evanescent waves which may have applications in nanoscale optomechanics, spintronics, and quantum information technology due to  the robust spin-directional coupling. Here we analyze a local spin angular momentum density of hybrid surface waves propagating along anisotropic hyperbolic metasurfaces. We reveal that, in contrast to bulk plane waves and conventional surface plasmons at isotropic interfaces, the spin of the hybrid surface waves can be engineered to have an arbitrary angle with the propagation direction. This property allows to tailor directivity of surface waves via the magnetic control of the spin projection of quantum emitters, and it can be useful for optically controlled spin transfer.
\end{abstract}

	
	\maketitle
	

\section{Introduction}

Metasurfaces are artificial two-dimensional nanostructured materials which exhibit new properties allowing to control and manage light propagation in an unusual way~\cite{yu2014flat}. Recently, there has been  significant interest in metasurfaces \cite{yu2015optical, Glybovski2016}, which is related to the interesting features and advantages that can be offered by these structures \cite{holloway2012overview}. {Keeping rich functionality of three-dimensional metamaterials, metasurfaces are simpler to fabricate, and they can be easily integrated into on-chip optical devices \cite{kildishev2013planar,Wu2014,Zhan2016}. Metasurfaces provide efficient beam shaping, phase and polarization control of light allowing to construct virtually arbitrary polarization vectors of the reflected or transmitted  waves ~\cite{Pfeiffer2013,Pors2013,Lin2014,Karimi2014,Arbabi2015,Veksler2015,Maguid2016}.

Another appealing feature of metasurfaces is that similar to bulk metamaterials providing efficient control over the bulk modes, metasurfaces provide unprecedented control over dispersion and polarization of surface waves~\cite{Rance2012,Shitrit2013,Martini2015,Li2014_1}. This idea was put forward in the seminal paper~\cite{Science_Engheta} in application to graphene metasurfaces, and it was  recently realized experimentally in the visible frequency range with the {plasmonic grating structure }~\cite{Lukin_2015}. Namely, a negative refraction of surface plasmon-polaritons has been demonstrated at the interface between silver and a {\it hyperbolic metasurface}~\cite{Lukin_2015}, i.e., a system characterized by the surface conductivity tensor with the principal components of different signs~\cite{gomez2015hyperbolic, yermakov2015hybrid, trushkov2015two, Hanson_2016}. It has been pointed out that the directivity of surface plasmons at the hyperbolic metasurfaces can be controlled with high flexibility allowing almost unidirectional propagation of surface waves excited by a point source.

The studies of surface waves have recently gained a new twist since it has been shown that they possess the unusual {\it transverse spin angular momentum}, perpendicular to their propagation direction~\cite{Bliokh2012,bliokh2014extraordinary}. Transverse spin is a generic feature of inhomogeneous light fields, and it has recently attracted considerable attention \cite{Bliokh20151,Aiello2015}. Importantly, a number of recent experiments \cite{Zayats2013,Petersen2014,Kuipers2015,Sollner2015} demonstrated that the transverse spin of evanescent waves provides a robust spin-direction coupling in a variety of optical systems \cite{bliokh2015quantum}. {Both transfer and control over the angular momentum of light at the nanoscale has a plethora of perspective applications in nanoscaled optomechanics~\cite{Gloppe2014,Neukirch2015} paving a way towards multidirectional mechanical control of the nanoobjects with light. Moreover, the coupling of the surface waves carrying the angular momentum to
  the magnetic solid state system leads to  the novel magneto-optical phenomena such as transversal magneto-optical Kerr effect~\cite{TMOKE1}, and it opens new perspectives for the efficient optical control over the spin currents in solid state systems, which is a subject of rapidly emerging field of \textit{spinoptronics}.} As such, it offers an efficient tool for spin-dependent control of light \cite{Bliokh2015SOI}.

\begin{figure}[htbp]
	\centering
	\includegraphics[width=1.0\columnwidth]{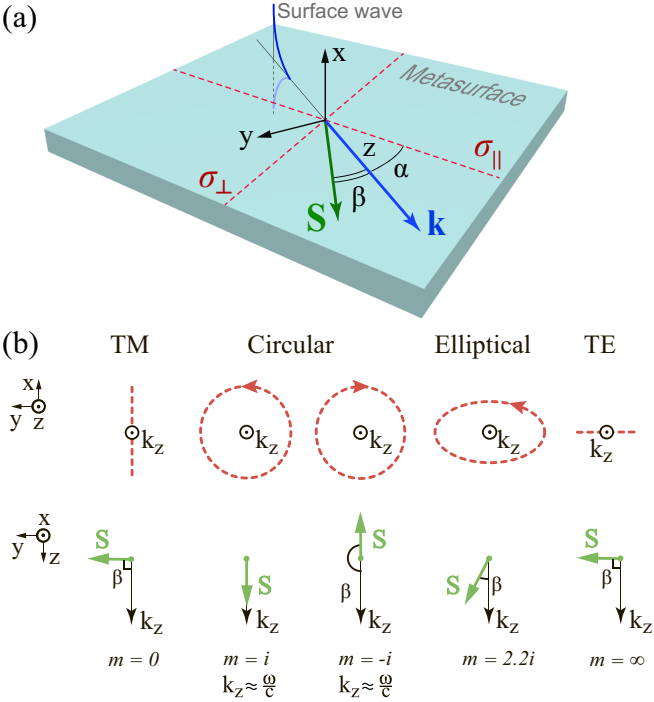}
	\caption{(a) Geometry of the structure. (b) Polarization of hyperbolic plasmons for different polarization parameters $m$ [see Eq.~\ref{SAM1}] with corresponding angles $\beta$ between plasmon propagation and spin angular momentum. The red arrow shows the direction of electric field vector rotation.}
	\label{fig:geometry_polarization}
\end{figure}

Electromagnetic eigenmodes of free space are {\it plane waves} whose spectrum (light cone) is double-degenerated with respect to polarization degrees of freedom. The spin angular momentum of a plane wave is always collinear to the propagation direction, and its projection on the wave vector (i.e., helicity) lies in the range $[-1,1]$. The helicity eigenmodes are left-hand and right-hand circularly-polarized waves.
Note that here and in what follows we consider pure momentum eigenmodes; superpositions of plane waves have more sophisticated spin properties \cite{Bliokh20151,Aiello2015}.

Electromagnetic {\it surface waves} can appear, e.g., at an interface between two isotropic media. In the generic case, the spectrum of surface waves is non-degenerate, and either linear TM or linear TE modes exists. For TM (TE) modes, electric (magnetic) field rotates in the plane perpendicular to the interface and containing the wave vector. Therefore, spin of surface waves lies in the plane of the interface and is directed perpendicular to the wave vector, in contrast to bulk plane waves \cite{bliokh2015quantum}. Thus, in isotropic media, the spin angular momentum can be either purely longitudinal for plane waves or purely transverse for surface waves. For nanophotonic applications, it would nevertheless be desirable to be able to tune spin direction and its absolute value continuously.

In this {work} we reveal that surface waves localized at {\it anisotropic} metasurfaces can serve as an instrument to fill this gap. We show that polarization of such surface waves can continuously change from linear TE or TM to elliptical, or circular (left of right) as a particular case. It provides controllable change of the angle between the spin and propagation direction in the range from 0 to $\pi$. These findings open new routes for both optomechanical manipulation of nanoobjects and for optical control over spin transport in semiconductor nanostructures coupled to hyperbolic metasurfaces.

\section{Dispersion of hybrid surface waves}

We consider a two-dimensional anisotropic structure shown in Fig.~\ref{fig:geometry_polarization}(a). Specific realizations include natural two-dimensional anisotropic materials such as hexagonal boron nitride~\cite{caldwell2014sub, li2015hyperbolic, dai2015subdiffractional}, plasmonic gratings of different geometries~\cite{liu2007far,ishii2013sub}, or patterned graphene nanostructures~\cite{iorsh2013hyperbolic,trushkov2015two}.  For microwaves, such anisotropic metasurfaces can be realized with the LC contours \cite{chshelokova2012hyperbolic}.

Within the local approximation the electromagnetic response of an anisotropic metasurface can be characterized by a surface conductivity tensor \cite{Glybovski2016}
	\begin{equation}
	\widehat\sigma_0=\left(
	\begin{matrix}
	\sigma_\bot & 0 \\
	0 &\sigma_\|
	\end{matrix}
	\right)
	\label{sigma_zero}.
	\end{equation}
We assume that conductivity tensor components have a resonance behaviour described by the Drude-Lorentz approximation:
	\begin{equation}
	\sigma_s(\omega)=A_s\frac{ic}{4\pi}\frac{\omega}{\omega^2-\Omega_{s}^2+i\gamma_s\omega}, \ \ s=\bot,\|.
	\label{sigma_disp}
	\end{equation}
Here $A_s$ is a constant, which depends on design, $\Omega_{s}$ is the resonant frequency, and $\gamma_s$ is the bandwidth of the resonance defined by losses. For the sake of simplicity, in what follows we neglect losses, $\gamma_s=0$, chose $A_s$ equal to $1$ and assume that $\Omega_\bot<\Omega_\|$. Three dispersion regimes of the metasurface can be distinguished depending on the signature of the conductivity tensor (1): (i) a capacitive regime at $\omega<\Omega_\bot$  when $\text{Im}(\sigma_\bot)<0$ and $\text{Im}(\sigma_\|)<0$; (ii) an inductive regime at $\omega>\Omega_\|$ when $\text{Im}(\sigma_\bot)>0$ and $\text{Im}(\sigma_\|)>0$; a hyperbolic regime at $\Omega_\bot<\omega<\Omega_\|$ when $\text{Im}(\sigma_\bot)\text{Im}(\sigma_\|)<0$.


The dispersion equation for surface waves propagating along the $z$-axis rotated by an angle $\alpha$ with respect to the $\|$-direction [see Fig.~\ref{fig:geometry_polarization}(a)] is given by~\cite{yermakov2015hybrid}:
\begin{align}
\label{dispersion}
&\left( \frac{\kappa}{k_0} - \frac{2\pi i}{c}\sigma_{yy}\right) \left( \frac{k_0}{\kappa} + \frac{2\pi i}{c}\sigma_{zz} \right) = \frac{4\pi^2}{c^2}\sigma_{yz}^2, &\\
& \ \ \begin{matrix}\sigma_{yy,zz}=\bar{\sigma}\mp \delta\sigma\cos (2\alpha),\\ \sigma_{yz}=\sigma_{zy}=\delta\sigma\sin(2\alpha). \end{matrix}&
\end{align}
Here we use the following notations: $\bar{\sigma}=(\sigma_{\perp}+\sigma_{\|})/2$, $\delta\sigma=(\sigma_{\|}-\sigma_{\perp})/2$, $k_0=\omega/c$, and $\kappa=\sqrt{k_z^2-k_0^2}$.

Angle $\alpha$ is the main parameter, which determines the propagation direction of the surface wave with respect to the anisotropy axis of the metasurface. It defines the relationship between the anisotropic properties of the metasurface and polarization (spin) properties of surface modes. When $\alpha=\pi n/2$ ($n$ is an integer number), the wave propagates along one of the anisotropy axes of the metasurface. In this case,} the right hand-side of Eq.~\eqref{dispersion} vanishes and the dispersion equation factorizes into two independent equations corresponding to the pure TE mode (left brackets) and the pure TM mode (right brackets). Straightforward analysis of Eq.~\eqref{dispersion} shows that (i) in the low-frequency (capacitive) regime the only TE mode is localized, (ii) in the high frequency (inductive) regime the only TM mode is localized, (iii) in the hyperbolic regime simultaneous propagation of both modes is possible.

For oblique propagation $\alpha\neq \pi n/2$ , the two linear polarizations get mixed and strictly speaking no specific linear polarization can be assigned to the eigenmodes of the structure. However, for the brevity, further on we will denote the upper frequency mode as {\it quasi-TM} surface wave and lower frequency as {\it quasi-TE} one.

The quasi-TE mode has no frequency cut-off, but has an angular-dependent resonant frequency $\omega^2_{\rm{r}}=\Omega_\|^2\sin^2(\alpha)+\Omega_\bot^2\cos^2(\alpha)$ for which $k_z(\omega_r)\rightarrow\infty$ [see Figs.~\ref{fig2}(a,c,e)]. The quasi-TM mode can propagate at an arbitrary high frequency but has the angular-dependent frequency cut-off $\omega^2_{\rm{c}}=\Omega_\|^2\cos^2(\alpha)+\Omega_\bot^2\sin^2(\alpha)$ [see Figs.~\ref{fig2}(a,c,e)].


\section{Spin angular momentum of hybrid surface waves}

{The spin angular momentum of a monochromatic electromagnetic field has intrinsic nature and is described by the local spin density. It should be noticed that total spin angular momentum of surface waves vanishes due to the symmetry of the problem. Total non-zero spin angular momentum can be revealed when permittivities of sub- and superstrate are different. However, in many practical problems local light-matter interactions (e.g., with atoms, nanoparticles, or quantum dots) are typically sensitive to the {\it local} spin density, in particular to the its {\it electric} part \cite{Zayats2013,Petersen2014,Kuipers2015,Sollner2015,Bliokh20151,Bliokh2014}, which usually has more sophisticated $x$-dependent properties than the total spin \cite{Bliokh20151}. Here and in what follows we will investigate the local spin density, which is the fundamental spin characteristic of surface modes, and we will denote it as \textit{spin angular momentum} for the brevity.
	
For surface waves, localized along the $x$-axis, we determine the local spin density normalized per ``one photon'' in units $\hbar=1$ as:
	\begin{equation}
	{\bf S} = \frac{\text{Im} {\bf [E^* \times E + H^* \times H ]} }{ W },
	\label{SAM}
	\end{equation}
where $W = |{\bf E}|^2 + |{\bf H}|^2$ characterizes the local energy density of the field. The dependence on the coordinates is excluded due to the normalization by the local energy density in Eq.~\eqref{SAM}.


In general case, we can express the electromagnetic fields through the polarization parameter $m$ in this way:
\begin{equation}
\begin{split}
& {\bf E} = \left( \pm 1, m\frac{k_0}{k_z}, -i\frac{\kappa}{k_z}   \right) e^{i k_z z - \kappa |x|}, \\
& {\bf H} = \left( -m, \pm \frac{k_0}{k_z}, \pm i m \frac{\kappa}{k_z}   \right) e^{i k_z z - \kappa |x|}.
\label{fields}
\end{split}
\end{equation}
Sign "$+$" corresponds to the upper half-space and "$-$" to the lower one.

To analyze the spin angular momentum of the surface waves supported by the anisotropic metasurface, we consider the upper half-space. Substituting electric and magnetic fields of surface modes at the anisotropic metasurface \eqref{fields} into Eq.~\eqref{SAM}, we arrive at the following expression: }
\begin{equation}
\mathbf{S}=\left(0, \frac{\kappa}{k_z}, \frac{2 \text{Im}(m)}{1 + |m|^2} \frac{k_0}{k_z}\right) \label{SAM1}~,~~
m=\frac{\frac{2\pi}{c}\sigma_{yz}}{1-i\frac{2\pi k_0}{\kappa c}\sigma_{yy}}.
\end{equation}
Here the complex polarization parameter $m$ is defined by the structure of the conductivity tensor in the surface waves, while for the plane waves $m$ represents just a ratio of the transverse electric-field components $E_y/E_x$ \cite{bliokh2014extraordinary,Bliokh20151}. Equation (\ref{SAM1}) is the central analytical result of our work, which describes {the local spin angular momentum density} of surface electromagnetic modes at anisotropic metasurfaces.

Importantly, in the isotropic case, $\sigma_{\perp}=\sigma_{\|}=\bar{\sigma}$ and $\delta\sigma=0$, when the conductivity tensor \eqref{sigma_zero} is scalar, only pure TM ($m$=0) and TE ($m=\infty$) surface modes exist. This case corresponds, for example, to {\it graphene plasmons} \cite{Mikhailov2007}. Equation (\ref{SAM1}) shows that graphene plasmons carry purely transverse spin  ${\bf S}=\left(0,\kappa/k_z,0\right)$ independent of the material properties.

For anisotropic metasurfaces, parameter $m$ can take on arbitrary values. Then, the surface modes are elliptically polarized, and their spin angular momentum is rotated by an angle $\beta$ with respect to the propagation direction, as shown in Fig.~\ref{fig:geometry_polarization}(b). Special cases of $m=\pm i$ correspond to right-hand and left-hand circularly polarized waves (with $\kappa \ll k_z \simeq k_0$ and almost longitudinal spin ${\bf S} \simeq (0,0,\pm 1) $), whereas the $m=0$ and $m=\infty$ cases correspond to TM or TE modes (with purely transverse spin ${\bf S}=\left(0,\kappa/k_z,0\right)$). All intermediate cases with arbitrary spin direction (within the metasurface plane) can be realized for the surface waves under consideration.



\begin{figure}[htbp]
		\centering
		\includegraphics[width = 0.96\columnwidth]{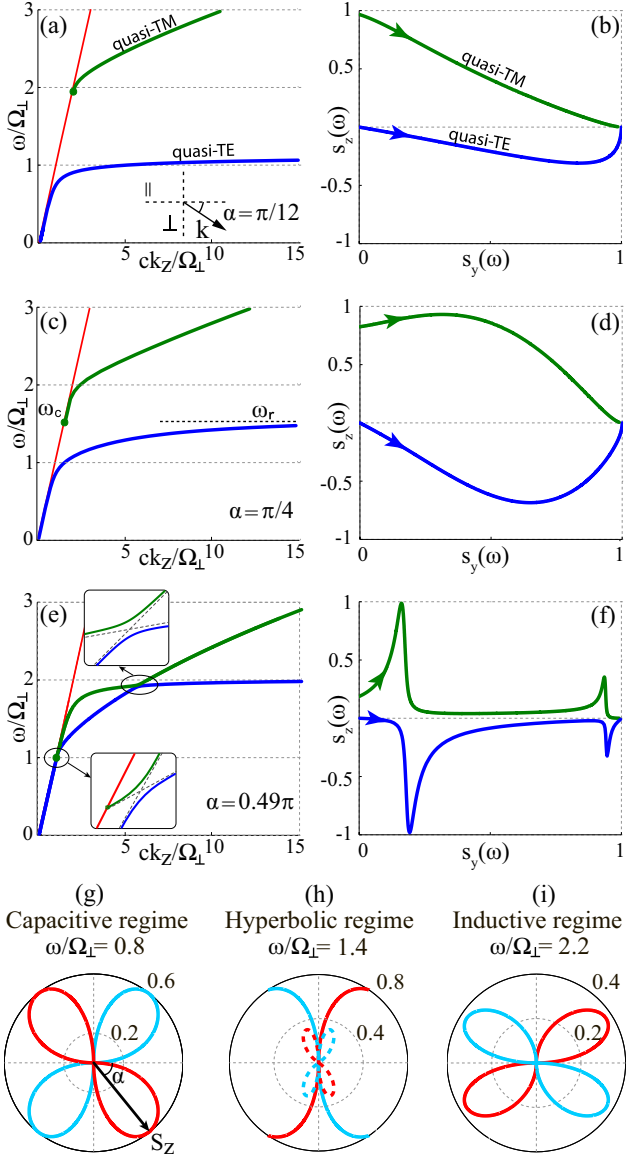}	
		\caption{Dispersion of the surface waves  (a,c,e) and corresponding parametric plots of spin angular momentum (b,d,f)  for quasi-TE (blue lines) and quasi-TM (green lines) modes for different propagation angles ($\alpha=\pi/12, \pi/4, 0.49\pi$). The red lines in figures (a,c,e) show the light line. The arrows in figures (b,d,f) correspond to increase of the frequency. (g-i) Dependence of longitudinal spin angular momentum $S_z$ on propagation direction $\alpha$ in polar coordinates for different frequencies $\omega/\Omega_\bot=0.8, 1.4, 2.2$. Red color corresponds to the positive sign and cyan to the negative one. In (h) solid lines corresponds to quasi-TM mode and dashed line to quasi-TE mode.}
		\label{fig2}
	\end{figure}

Pure TE and TM surface modes appear for $\alpha=\pi n/2$. For the oblique propagation $\alpha \neq \pi n/2$, quasi-TE and quasi-TM modes are elliptically polarized and have non-zero longitudinal and transverse spin components. The evolution of these components $S_z$ and $S_y$ along the dispersion curves of the quasi-TE and quasi-TM surface waves for different propagation angles ($\alpha=\pi/12, \pi/4, 0.49\pi$) are shown in Figs.~\ref{fig2}(b,d,f). The corresponding dispersions of the surface waves are shown in Figs.~\ref{fig2}(a,c,e). The longitudinal spin component of the quasi-TM mode can change from 1 [see Fig.~\ref{fig2}(b)] in the vicinity of frequency cut-off to 0 at high frequencies, whereas the longitudinal spin of the quasi-TE approaches zero both for the small and large $k_z$.

Comparison of Figs.~\ref{fig2}(b) and ~\ref{fig2}(d) shows that the hybridization is stronger when the propagation direction becomes farther from the the principal axes. So, at $\alpha=\pi/4$, it results in relatively-large longitudinal spin $S_z$ for the quasi-TE mode. For the quasi-TM mode $S_z$ reaches 1 at some finite $k_z$.

In the hyperbolic regime of the metasurface, simultaneous propagation of two modes is possible. For $\alpha=\pm\pi/2$, the dispersion curves of the TE and TM modes with purely transverse spin crosses, i.e.  an {\it accidental degeneracy} takes place.
Small deviation of $\alpha$ from $\pm\pi/2$ lifts the degeneracies and anticrossing gaps open [see Fig.~\ref{fig2}(e)]. The anticrossings signify the {\it hybridization} of the TE and TM modes. This brings about sharp resonances in the longitudinal spin components [Fig.~\ref{fig2}(f)]. In these resonances, hybridized surface eigenmodes acquire circularly polarizations and can posses nearly longitudinal spin. This is in contrast to the usual surface waves with purely transverse spin.

\begin{figure}[htbp]
	\centering
	\includegraphics[width=0.9\linewidth]{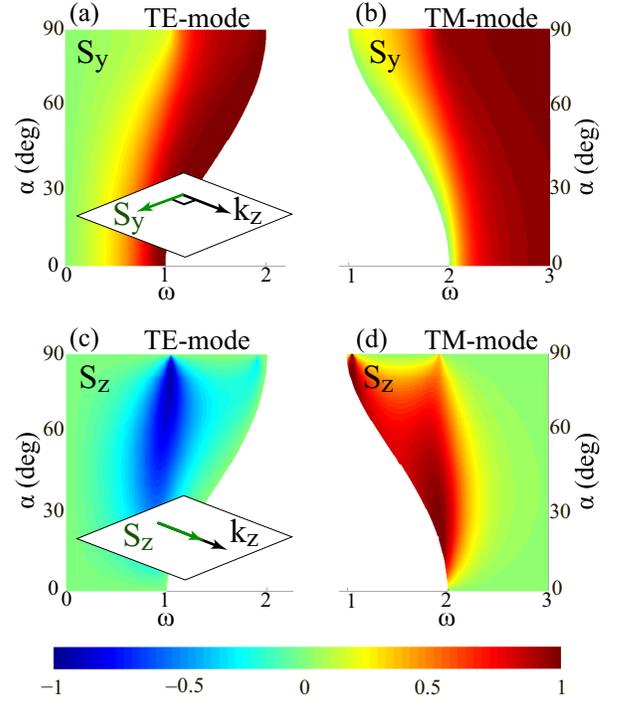}
	\caption{Spin angular momentum isocurves as functions of frequency $\omega$ and propagation angle $\alpha$ for transverse component of spin angular momentum for TE (a) and for TM (b) modes, longitudinal component of spin angular momentum for TE (c) and for TM (d) modes.}
	\label{fig3}
\end{figure}

Notably, the quasi-TE and quasi-TM modes always have transverse spins of the same sign and longitudinal spins of opposite signs. The dependances of  $S_z$ on  $\alpha$ for three different dispersion regimes of metasurface is shown in Figs.~\ref{fig2}(g,h,i). For the quasi-TE mode, $S_z < 0$ when $\alpha$ lies in the first and third quadrants and $S_z > 0$ in the second and forth quadrants. For the quasi-TM modes, the situation is reversed. Note that for the quasi-TM mode in the hyperbolic regime the contours are open [Fig.~\ref{fig2}(h)]. This is due to the hyperbolic shape of the equal frequency contours, which forbids the propagation of the surface wave in certain range of angles.

Comprehensive information about angular and frequency dependances of the transverse ($S_y$) and longitudinal ($S_z$) spin components for quasi-TE and quasi-TM modes is provided in Fig.~\ref{fig3}. As discussed previously, the transverse spin increases monotonically with frequency for a fixed propagation direction for the both modes. At the same time, the longitudinal spin behaviour depends crucially on the propagation direction $\alpha$. When $\alpha \rightarrow \pi/2$ one can see two resonances both for quasi-TM and quasi-TE modes, which are also seen in Fig.~\ref{fig2}(f).

Finally, it should be emphasized that there are two mechanisms of the hybridization of surface TE and TM modes, which determine their spin angular momenta: (i) smooth anisotropy-induced hybridization of modes into quasi-TE and quasi-TM eigenmodes and (ii) resonant hybridization due to an accidental degeneracy of the eigenmodes in the hyperbolic regime. We notice also that two-dimensional control of the surface-wave spin can be extented to the full three-dimensional case by considering {\it chiral metasurfaces}. In this case, one can expect that eigenmodes will have also the vertical spin component.

\section{Conclusion}

We have analyzed the spin angular momentum of surface waves localized at anisotropic metasurfaces. We have shown that hyperbolic metasurfaces allow flexible control of both the longitudinal and transverse components of the spin angular momentum of surface waves. This finding is in a sharp contrast to the properties of conventional surface waves localized at interfaces of isotropic materials, which carry purely transverse spin. Two-dimensional tunability of optical spin at anisotropic interfaces can enrich considerably various spin-orbit interaction phenomena, which currently attract enormous 
attention in nanophotonics and near-field optics \cite{Bliokh2015SOI,Zayats2013,Petersen2014,Kuipers2015,Sollner2015,bliokh2015quantum}.

\begin{acknowledgments}
 {This work was supported by RFBR (16-37-60064, 14-02-01223), by the President of Russian Federation (MK-6462.2016.2), the Federal Programme on Support of Leading Scientific Schools (NSh-5062.2014.2), the program of Fundamental Research in Nanotechnology and Nanomaterials of the Russian Academy of Science, and the Australian Research Council. Numerical simulations and optical experiments have been supported by the Russian Science Foundation (Grant 15-12-20028). O.Y. acknowledges a support from the Dynasty Foundation.}
\end{acknowledgments}
	
\bibliography{refer}

\begin{thebibliography}{46}%
\makeatletter
\providecommand \@ifxundefined [1]{%
 \@ifx{#1\undefined}
}%
\providecommand \@ifnum [1]{%
 \ifnum #1\expandafter \@firstoftwo
 \else \expandafter \@secondoftwo
 \fi
}%
\providecommand \@ifx [1]{%
 \ifx #1\expandafter \@firstoftwo
 \else \expandafter \@secondoftwo
 \fi
}%
\providecommand \natexlab [1]{#1}%
\providecommand \enquote  [1]{``#1''}%
\providecommand \bibnamefont  [1]{#1}%
\providecommand \bibfnamefont [1]{#1}%
\providecommand \citenamefont [1]{#1}%
\providecommand \href@noop [0]{\@secondoftwo}%
\providecommand \href [0]{\begingroup \@sanitize@url \@href}%
\providecommand \@href[1]{\@@startlink{#1}\@@href}%
\providecommand \@@href[1]{\endgroup#1\@@endlink}%
\providecommand \@sanitize@url [0]{\catcode `\\12\catcode `\$12\catcode
  `\&12\catcode `\#12\catcode `\^12\catcode `\_12\catcode `\%12\relax}%
\providecommand \@@startlink[1]{}%
\providecommand \@@endlink[0]{}%
\providecommand \url  [0]{\begingroup\@sanitize@url \@url }%
\providecommand \@url [1]{\endgroup\@href {#1}{\urlprefix }}%
\providecommand \urlprefix  [0]{URL }%
\providecommand \Eprint [0]{\href }%
\providecommand \doibase [0]{http://dx.doi.org/}%
\providecommand \selectlanguage [0]{\@gobble}%
\providecommand \bibinfo  [0]{\@secondoftwo}%
\providecommand \bibfield  [0]{\@secondoftwo}%
\providecommand \translation [1]{[#1]}%
\providecommand \BibitemOpen [0]{}%
\providecommand \bibitemStop [0]{}%
\providecommand \bibitemNoStop [0]{.\EOS\space}%
\providecommand \EOS [0]{\spacefactor3000\relax}%
\providecommand \BibitemShut  [1]{\csname bibitem#1\endcsname}%
\let\auto@bib@innerbib\@empty
\bibitem [{\citenamefont {Yu}\ and\ \citenamefont
  {Capasso}(2014)}]{yu2014flat}%
  \BibitemOpen
  \bibfield  {author} {\bibinfo {author} {\bibfnamefont {N.}~\bibnamefont
  {Yu}}\ and\ \bibinfo {author} {\bibfnamefont {F.}~\bibnamefont {Capasso}},\
  }\href@noop {} {\bibfield  {journal} {\bibinfo  {journal} {Nat. Mater.}\
  }\textbf {\bibinfo {volume} {13}},\ \bibinfo {pages} {139} (\bibinfo {year}
  {2014})}\BibitemShut {NoStop}%
\bibitem [{\citenamefont {Yu}\ and\ \citenamefont
  {Capasso}(2015)}]{yu2015optical}%
  \BibitemOpen
  \bibfield  {author} {\bibinfo {author} {\bibfnamefont {N.}~\bibnamefont
  {Yu}}\ and\ \bibinfo {author} {\bibfnamefont {F.}~\bibnamefont {Capasso}},\
  }\href@noop {} {\bibfield  {journal} {\bibinfo  {journal} {J. Lightwave
  Technol.}\ }\textbf {\bibinfo {volume} {33}},\ \bibinfo {pages} {2344}
  (\bibinfo {year} {2015})}\BibitemShut {NoStop}%
\bibitem [{\citenamefont {Glybovski}\ \emph {et~al.}(2016)\citenamefont
  {Glybovski}, \citenamefont {Tretyakov}, \citenamefont {Belov}, \citenamefont
  {Kivshar},\ and\ \citenamefont {Simovski}}]{Glybovski2016}%
  \BibitemOpen
  \bibfield  {author} {\bibinfo {author} {\bibfnamefont {S.~B.}\ \bibnamefont
  {Glybovski}}, \bibinfo {author} {\bibfnamefont {S.~A.}\ \bibnamefont
  {Tretyakov}}, \bibinfo {author} {\bibfnamefont {P.~A.}\ \bibnamefont
  {Belov}}, \bibinfo {author} {\bibfnamefont {Y.~S.}\ \bibnamefont {Kivshar}},
  \ and\ \bibinfo {author} {\bibfnamefont {C.~R.}\ \bibnamefont {Simovski}},\
  }\href {\doibase http://dx.doi.org/10.1016/j.physrep.2016.04.004} {\bibfield
  {journal} {\bibinfo  {journal} {Phys. Rep.}\ }\textbf {\bibinfo {volume}
  {634}},\ \bibinfo {pages} {1 } (\bibinfo {year} {2016})}\BibitemShut
  {NoStop}%
\bibitem [{\citenamefont {Holloway}\ \emph {et~al.}(2012)\citenamefont
  {Holloway}, \citenamefont {Kuester}, \citenamefont {Gordon}, \citenamefont
  {Hara}, \citenamefont {Booth},\ and\ \citenamefont
  {Smith}}]{holloway2012overview}%
  \BibitemOpen
  \bibfield  {author} {\bibinfo {author} {\bibfnamefont {C.~L.}\ \bibnamefont
  {Holloway}}, \bibinfo {author} {\bibfnamefont {E.~F.}\ \bibnamefont
  {Kuester}}, \bibinfo {author} {\bibfnamefont {J.}~\bibnamefont {Gordon}},
  \bibinfo {author} {\bibfnamefont {J.~O.}\ \bibnamefont {Hara}}, \bibinfo
  {author} {\bibfnamefont {J.}~\bibnamefont {Booth}}, \ and\ \bibinfo {author}
  {\bibfnamefont {D.~R.}\ \bibnamefont {Smith}},\ }\href@noop {} {\bibfield
  {journal} {\bibinfo  {journal} {IEEE Antenn. Propag. M.}\ }\textbf {\bibinfo
  {volume} {54}},\ \bibinfo {pages} {10} (\bibinfo {year} {2012})}\BibitemShut
  {NoStop}%
\bibitem [{\citenamefont {Kildishev}\ \emph {et~al.}(2013)\citenamefont
  {Kildishev}, \citenamefont {Boltasseva},\ and\ \citenamefont
  {Shalaev}}]{kildishev2013planar}%
  \BibitemOpen
  \bibfield  {author} {\bibinfo {author} {\bibfnamefont {A.~V.}\ \bibnamefont
  {Kildishev}}, \bibinfo {author} {\bibfnamefont {A.}~\bibnamefont
  {Boltasseva}}, \ and\ \bibinfo {author} {\bibfnamefont {V.~M.}\ \bibnamefont
  {Shalaev}},\ }\href@noop {} {\bibfield  {journal} {\bibinfo  {journal}
  {Science}\ }\textbf {\bibinfo {volume} {339}},\ \bibinfo {pages} {1232009}
  (\bibinfo {year} {2013})}\BibitemShut {NoStop}%
\bibitem [{\citenamefont {Wu}\ \emph {et~al.}(2014)\citenamefont {Wu},
  \citenamefont {Arju}, \citenamefont {Kelp}, \citenamefont {Fan},
  \citenamefont {Dominguez}, \citenamefont {Gonzales}, \citenamefont {Tutuc},
  \citenamefont {Brener},\ and\ \citenamefont {Shvets}}]{Wu2014}%
  \BibitemOpen
  \bibfield  {author} {\bibinfo {author} {\bibfnamefont {C.}~\bibnamefont
  {Wu}}, \bibinfo {author} {\bibfnamefont {N.}~\bibnamefont {Arju}}, \bibinfo
  {author} {\bibfnamefont {G.}~\bibnamefont {Kelp}}, \bibinfo {author}
  {\bibfnamefont {J.~A.}\ \bibnamefont {Fan}}, \bibinfo {author} {\bibfnamefont
  {J.}~\bibnamefont {Dominguez}}, \bibinfo {author} {\bibfnamefont
  {E.}~\bibnamefont {Gonzales}}, \bibinfo {author} {\bibfnamefont
  {E.}~\bibnamefont {Tutuc}}, \bibinfo {author} {\bibfnamefont
  {I.}~\bibnamefont {Brener}}, \ and\ \bibinfo {author} {\bibfnamefont
  {G.}~\bibnamefont {Shvets}},\ }\href@noop {} {\bibfield  {journal} {\bibinfo
  {journal} {Nat. Commun.}\ }\textbf {\bibinfo {volume} {5}},\ \bibinfo {pages}
  {3892} (\bibinfo {year} {2014})}\BibitemShut {NoStop}%
\bibitem [{\citenamefont {Zhan}\ \emph {et~al.}(2016)\citenamefont {Zhan},
  \citenamefont {Colburn}, \citenamefont {Trivedi}, \citenamefont {Dodson},\
  and\ \citenamefont {Majumdar}}]{Zhan2016}%
  \BibitemOpen
  \bibfield  {author} {\bibinfo {author} {\bibfnamefont {A.}~\bibnamefont
  {Zhan}}, \bibinfo {author} {\bibfnamefont {S.}~\bibnamefont {Colburn}},
  \bibinfo {author} {\bibfnamefont {R.}~\bibnamefont {Trivedi}}, \bibinfo
  {author} {\bibfnamefont {C.}~\bibnamefont {Dodson}}, \ and\ \bibinfo {author}
  {\bibfnamefont {A.}~\bibnamefont {Majumdar}},\ }\href@noop {} {\bibfield
  {journal} {\bibinfo  {journal} {ACS Photonics}\ }\textbf {\bibinfo {volume}
  {3}},\ \bibinfo {pages} {209} (\bibinfo {year} {2016})}\BibitemShut {NoStop}%
\bibitem [{\citenamefont {Pfeiffer}\ and\ \citenamefont
  {Grbic}(2013)}]{Pfeiffer2013}%
  \BibitemOpen
  \bibfield  {author} {\bibinfo {author} {\bibfnamefont {C.}~\bibnamefont
  {Pfeiffer}}\ and\ \bibinfo {author} {\bibfnamefont {A.}~\bibnamefont
  {Grbic}},\ }\href@noop {} {\bibfield  {journal} {\bibinfo  {journal} {Phys.
  Rev. Lett.}\ }\textbf {\bibinfo {volume} {110}},\ \bibinfo {pages} {197401}
  (\bibinfo {year} {2013})}\BibitemShut {NoStop}%
\bibitem [{\citenamefont {Pors}\ and\ \citenamefont
  {Bozhevolnyi}(2013)}]{Pors2013}%
  \BibitemOpen
  \bibfield  {author} {\bibinfo {author} {\bibfnamefont {A.}~\bibnamefont
  {Pors}}\ and\ \bibinfo {author} {\bibfnamefont {S.~I.}\ \bibnamefont
  {Bozhevolnyi}},\ }\href@noop {} {\bibfield  {journal} {\bibinfo  {journal}
  {Opt. Express}\ }\textbf {\bibinfo {volume} {21}},\ \bibinfo {pages} {27438}
  (\bibinfo {year} {2013})}\BibitemShut {NoStop}%
\bibitem [{\citenamefont {Lin}\ \emph {et~al.}(2014)\citenamefont {Lin},
  \citenamefont {Fan}, \citenamefont {Hasman},\ and\ \citenamefont
  {Brongersma}}]{Lin2014}%
  \BibitemOpen
  \bibfield  {author} {\bibinfo {author} {\bibfnamefont {D.}~\bibnamefont
  {Lin}}, \bibinfo {author} {\bibfnamefont {P.}~\bibnamefont {Fan}}, \bibinfo
  {author} {\bibfnamefont {E.}~\bibnamefont {Hasman}}, \ and\ \bibinfo {author}
  {\bibfnamefont {M.~L.}\ \bibnamefont {Brongersma}},\ }\href@noop {}
  {\bibfield  {journal} {\bibinfo  {journal} {Science}\ }\textbf {\bibinfo
  {volume} {345}},\ \bibinfo {pages} {298} (\bibinfo {year}
  {2014})}\BibitemShut {NoStop}%
\bibitem [{\citenamefont {Karimi}\ \emph {et~al.}(2014)\citenamefont {Karimi},
  \citenamefont {Schulz}, \citenamefont {De~Leon}, \citenamefont {Qassim},
  \citenamefont {Upham},\ and\ \citenamefont {Boyd}}]{Karimi2014}%
  \BibitemOpen
  \bibfield  {author} {\bibinfo {author} {\bibfnamefont {E.}~\bibnamefont
  {Karimi}}, \bibinfo {author} {\bibfnamefont {S.~A.}\ \bibnamefont {Schulz}},
  \bibinfo {author} {\bibfnamefont {I.}~\bibnamefont {De~Leon}}, \bibinfo
  {author} {\bibfnamefont {H.}~\bibnamefont {Qassim}}, \bibinfo {author}
  {\bibfnamefont {J.}~\bibnamefont {Upham}}, \ and\ \bibinfo {author}
  {\bibfnamefont {R.~W.}\ \bibnamefont {Boyd}},\ }\href@noop {} {\bibfield
  {journal} {\bibinfo  {journal} {Light Sci. Appl.}\ }\textbf {\bibinfo
  {volume} {3}},\ \bibinfo {pages} {e167} (\bibinfo {year} {2014})}\BibitemShut
  {NoStop}%
\bibitem [{\citenamefont {Arbabi}\ \emph {et~al.}(2015)\citenamefont {Arbabi},
  \citenamefont {Horie}, \citenamefont {Bagheri},\ and\ \citenamefont
  {Faraon}}]{Arbabi2015}%
  \BibitemOpen
  \bibfield  {author} {\bibinfo {author} {\bibfnamefont {A.}~\bibnamefont
  {Arbabi}}, \bibinfo {author} {\bibfnamefont {Y.}~\bibnamefont {Horie}},
  \bibinfo {author} {\bibfnamefont {M.}~\bibnamefont {Bagheri}}, \ and\
  \bibinfo {author} {\bibfnamefont {A.}~\bibnamefont {Faraon}},\ }\href@noop {}
  {\bibfield  {journal} {\bibinfo  {journal} {Nat. Nanotechnol.}\ }\textbf
  {\bibinfo {volume} {10}},\ \bibinfo {pages} {937} (\bibinfo {year}
  {2015})}\BibitemShut {NoStop}%
\bibitem [{\citenamefont {Veksler}\ \emph {et~al.}(2015)\citenamefont
  {Veksler}, \citenamefont {Maguid}, \citenamefont {Shitrit}, \citenamefont
  {Ozeri}, \citenamefont {Kleiner},\ and\ \citenamefont
  {Hasman}}]{Veksler2015}%
  \BibitemOpen
  \bibfield  {author} {\bibinfo {author} {\bibfnamefont {D.}~\bibnamefont
  {Veksler}}, \bibinfo {author} {\bibfnamefont {E.}~\bibnamefont {Maguid}},
  \bibinfo {author} {\bibfnamefont {N.}~\bibnamefont {Shitrit}}, \bibinfo
  {author} {\bibfnamefont {D.}~\bibnamefont {Ozeri}}, \bibinfo {author}
  {\bibfnamefont {V.}~\bibnamefont {Kleiner}}, \ and\ \bibinfo {author}
  {\bibfnamefont {E.}~\bibnamefont {Hasman}},\ }\href@noop {} {\bibfield
  {journal} {\bibinfo  {journal} {ACS Photonics}\ }\textbf {\bibinfo {volume}
  {2}},\ \bibinfo {pages} {661–} (\bibinfo {year} {2015})}\BibitemShut
  {NoStop}%
\bibitem [{\citenamefont {Maguid}\ \emph {et~al.}(2016)\citenamefont {Maguid},
  \citenamefont {Yulevich}, \citenamefont {Veksler}, \citenamefont {Kleiner},
  \citenamefont {Brongersma},\ and\ \citenamefont {Hasman}}]{Maguid2016}%
  \BibitemOpen
  \bibfield  {author} {\bibinfo {author} {\bibfnamefont {E.}~\bibnamefont
  {Maguid}}, \bibinfo {author} {\bibfnamefont {I.}~\bibnamefont {Yulevich}},
  \bibinfo {author} {\bibfnamefont {D.}~\bibnamefont {Veksler}}, \bibinfo
  {author} {\bibfnamefont {V.}~\bibnamefont {Kleiner}}, \bibinfo {author}
  {\bibfnamefont {M.}~\bibnamefont {Brongersma}}, \ and\ \bibinfo {author}
  {\bibfnamefont {E.}~\bibnamefont {Hasman}},\ }\href@noop {} {\bibfield
  {journal} {\bibinfo  {journal} {Science}\ ,\ \bibinfo {pages}
  {10.1126/science.aaf3417}} (\bibinfo {year} {2016})}\BibitemShut {NoStop}%
\bibitem [{\citenamefont {Rance}\ \emph {et~al.}(2012)\citenamefont {Rance},
  \citenamefont {Constant}, \citenamefont {Hibbins},\ and\ \citenamefont
  {Sambles}}]{Rance2012}%
  \BibitemOpen
  \bibfield  {author} {\bibinfo {author} {\bibfnamefont {H.~J.}\ \bibnamefont
  {Rance}}, \bibinfo {author} {\bibfnamefont {T.~J.}\ \bibnamefont {Constant}},
  \bibinfo {author} {\bibfnamefont {A.~P.}\ \bibnamefont {Hibbins}}, \ and\
  \bibinfo {author} {\bibfnamefont {J.~R.}\ \bibnamefont {Sambles}},\
  }\href@noop {} {\bibfield  {journal} {\bibinfo  {journal} {Phys. Rev. B}\
  }\textbf {\bibinfo {volume} {86}},\ \bibinfo {pages} {125144} (\bibinfo
  {year} {2012})}\BibitemShut {NoStop}%
\bibitem [{\citenamefont {Shitrit}\ \emph {et~al.}(2013)\citenamefont
  {Shitrit}, \citenamefont {Yulevich}, \citenamefont {Maguid}, \citenamefont
  {Ozeri}, \citenamefont {Veksler}, \citenamefont {Kleiner},\ and\
  \citenamefont {Hasman}}]{Shitrit2013}%
  \BibitemOpen
  \bibfield  {author} {\bibinfo {author} {\bibfnamefont {N.}~\bibnamefont
  {Shitrit}}, \bibinfo {author} {\bibfnamefont {I.}~\bibnamefont {Yulevich}},
  \bibinfo {author} {\bibfnamefont {E.}~\bibnamefont {Maguid}}, \bibinfo
  {author} {\bibfnamefont {D.}~\bibnamefont {Ozeri}}, \bibinfo {author}
  {\bibfnamefont {D.}~\bibnamefont {Veksler}}, \bibinfo {author} {\bibfnamefont
  {V.}~\bibnamefont {Kleiner}}, \ and\ \bibinfo {author} {\bibfnamefont
  {E.}~\bibnamefont {Hasman}},\ }\href@noop {} {\bibfield  {journal} {\bibinfo
  {journal} {Science}\ }\textbf {\bibinfo {volume} {340}},\ \bibinfo {pages}
  {724–} (\bibinfo {year} {2013})}\BibitemShut {NoStop}%
\bibitem [{\citenamefont {Martini}\ \emph {et~al.}(2015)\citenamefont
  {Martini}, \citenamefont {Mencagli},\ and\ \citenamefont
  {Maci}}]{Martini2015}%
  \BibitemOpen
  \bibfield  {author} {\bibinfo {author} {\bibfnamefont {E.}~\bibnamefont
  {Martini}}, \bibinfo {author} {\bibfnamefont {M.}~\bibnamefont {Mencagli}}, \
  and\ \bibinfo {author} {\bibfnamefont {S.}~\bibnamefont {Maci}},\ }\href@noop
  {} {\bibfield  {journal} {\bibinfo  {journal} {Phil. T. R. Soc. A}\ }\textbf
  {\bibinfo {volume} {373}},\ \bibinfo {pages} {20140355} (\bibinfo {year}
  {2015})}\BibitemShut {NoStop}%
\bibitem [{\citenamefont {Li}\ \emph {et~al.}(2014)\citenamefont {Li},
  \citenamefont {Cai}, \citenamefont {Wan},\ and\ \citenamefont
  {Cui}}]{Li2014_1}%
  \BibitemOpen
  \bibfield  {author} {\bibinfo {author} {\bibfnamefont {Y.~B.}\ \bibnamefont
  {Li}}, \bibinfo {author} {\bibfnamefont {B.~G.}\ \bibnamefont {Cai}},
  \bibinfo {author} {\bibfnamefont {X.}~\bibnamefont {Wan}}, \ and\ \bibinfo
  {author} {\bibfnamefont {T.~J.}\ \bibnamefont {Cui}},\ }\href@noop {}
  {\bibfield  {journal} {\bibinfo  {journal} {Opt. Lett.}\ }\textbf {\bibinfo
  {volume} {39}},\ \bibinfo {pages} {5888} (\bibinfo {year}
  {2014})}\BibitemShut {NoStop}%
\bibitem [{\citenamefont {Vakil}\ and\ \citenamefont
  {Engheta}(2007)}]{Science_Engheta}%
  \BibitemOpen
  \bibfield  {author} {\bibinfo {author} {\bibfnamefont {A.}~\bibnamefont
  {Vakil}}\ and\ \bibinfo {author} {\bibfnamefont {N.}~\bibnamefont
  {Engheta}},\ }\href@noop {} {\bibfield  {journal} {\bibinfo  {journal}
  {Science}\ }\textbf {\bibinfo {volume} {332}},\ \bibinfo {pages} {1291}
  (\bibinfo {year} {2007})}\BibitemShut {NoStop}%
\bibitem [{\citenamefont {High}\ \emph {et~al.}(2015)\citenamefont {High},
  \citenamefont {R.~C.~Devlin}, \citenamefont {Polking}, \citenamefont {Wild},
  \citenamefont {Perczel}, \citenamefont {de~Leon}, \citenamefont {Lukin},\
  and\ \citenamefont {Park}}]{Lukin_2015}%
  \BibitemOpen
  \bibfield  {author} {\bibinfo {author} {\bibfnamefont {A.}~\bibnamefont
  {High}}, \bibinfo {author} {\bibfnamefont {A.~D.}\ \bibnamefont
  {R.~C.~Devlin}}, \bibinfo {author} {\bibfnamefont {M.}~\bibnamefont
  {Polking}}, \bibinfo {author} {\bibfnamefont {D.}~\bibnamefont {Wild}},
  \bibinfo {author} {\bibfnamefont {J.}~\bibnamefont {Perczel}}, \bibinfo
  {author} {\bibfnamefont {N.~P.}\ \bibnamefont {de~Leon}}, \bibinfo {author}
  {\bibfnamefont {M.}~\bibnamefont {Lukin}}, \ and\ \bibinfo {author}
  {\bibfnamefont {H.}~\bibnamefont {Park}},\ }\href@noop {} {\bibfield
  {journal} {\bibinfo  {journal} {Nature}\ }\textbf {\bibinfo {volume} {522}},\
  \bibinfo {pages} {192} (\bibinfo {year} {2015})}\BibitemShut {NoStop}%
\bibitem [{\citenamefont {Gomez-Diaz}\ \emph {et~al.}(2015)\citenamefont
  {Gomez-Diaz}, \citenamefont {Tymchenko},\ and\ \citenamefont
  {Al{\`u}}}]{gomez2015hyperbolic}%
  \BibitemOpen
  \bibfield  {author} {\bibinfo {author} {\bibfnamefont {J.~S.}\ \bibnamefont
  {Gomez-Diaz}}, \bibinfo {author} {\bibfnamefont {M.}~\bibnamefont
  {Tymchenko}}, \ and\ \bibinfo {author} {\bibfnamefont {A.}~\bibnamefont
  {Al{\`u}}},\ }\href@noop {} {\bibfield  {journal} {\bibinfo  {journal} {Phys.
  Rev. Lett.}\ }\textbf {\bibinfo {volume} {114}},\ \bibinfo {pages} {233901}
  (\bibinfo {year} {2015})}\BibitemShut {NoStop}%
\bibitem [{\citenamefont {Yermakov}\ \emph {et~al.}(2015)\citenamefont
  {Yermakov}, \citenamefont {Ovcharenko}, \citenamefont {Song}, \citenamefont
  {Bogdanov}, \citenamefont {Iorsh},\ and\ \citenamefont
  {Kivshar}}]{yermakov2015hybrid}%
  \BibitemOpen
  \bibfield  {author} {\bibinfo {author} {\bibfnamefont {O.~Y.}\ \bibnamefont
  {Yermakov}}, \bibinfo {author} {\bibfnamefont {A.~I.}\ \bibnamefont
  {Ovcharenko}}, \bibinfo {author} {\bibfnamefont {M.}~\bibnamefont {Song}},
  \bibinfo {author} {\bibfnamefont {A.~A.}\ \bibnamefont {Bogdanov}}, \bibinfo
  {author} {\bibfnamefont {I.~V.}\ \bibnamefont {Iorsh}}, \ and\ \bibinfo
  {author} {\bibfnamefont {Y.~S.}\ \bibnamefont {Kivshar}},\ }\href {\doibase
  10.1103/PhysRevB.91.235423} {\bibfield  {journal} {\bibinfo  {journal} {Phys.
  Rev. B}\ }\textbf {\bibinfo {volume} {91}},\ \bibinfo {pages} {235423}
  (\bibinfo {year} {2015})}\BibitemShut {NoStop}%
\bibitem [{\citenamefont {Trushkov}\ and\ \citenamefont
  {Iorsh}(2015)}]{trushkov2015two}%
  \BibitemOpen
  \bibfield  {author} {\bibinfo {author} {\bibfnamefont {I.}~\bibnamefont
  {Trushkov}}\ and\ \bibinfo {author} {\bibfnamefont {I.}~\bibnamefont
  {Iorsh}},\ }\href@noop {} {\bibfield  {journal} {\bibinfo  {journal} {Phys.
  Rev. B}\ }\textbf {\bibinfo {volume} {92}},\ \bibinfo {pages} {045305}
  (\bibinfo {year} {2015})}\BibitemShut {NoStop}%
\bibitem [{\citenamefont {Nemilentsau}\ \emph {et~al.}(2016)\citenamefont
  {Nemilentsau}, \citenamefont {Low},\ and\ \citenamefont
  {Hanson}}]{Hanson_2016}%
  \BibitemOpen
  \bibfield  {author} {\bibinfo {author} {\bibfnamefont {A.}~\bibnamefont
  {Nemilentsau}}, \bibinfo {author} {\bibfnamefont {T.}~\bibnamefont {Low}}, \
  and\ \bibinfo {author} {\bibfnamefont {G.}~\bibnamefont {Hanson}},\ }\href
  {\doibase 10.1103/PhysRevLett.116.066804} {\bibfield  {journal} {\bibinfo
  {journal} {Phys. Rev. Lett.}\ }\textbf {\bibinfo {volume} {116}},\ \bibinfo
  {pages} {066804} (\bibinfo {year} {2016})}\BibitemShut {NoStop}%
\bibitem [{\citenamefont {Bliokh}\ and\ \citenamefont
  {Nori}(2012)}]{Bliokh2012}%
  \BibitemOpen
  \bibfield  {author} {\bibinfo {author} {\bibfnamefont {K.}~\bibnamefont
  {Bliokh}}\ and\ \bibinfo {author} {\bibfnamefont {F.}~\bibnamefont {Nori}},\
  }\href@noop {} {\bibfield  {journal} {\bibinfo  {journal} {Phys. Rev. A}\
  }\textbf {\bibinfo {volume} {85}},\ \bibinfo {pages} {061801(R)} (\bibinfo
  {year} {2012})}\BibitemShut {NoStop}%
\bibitem [{\citenamefont {Bliokh}\ \emph
  {et~al.}(2014{\natexlab{a}})\citenamefont {Bliokh}, \citenamefont
  {Bekshaev},\ and\ \citenamefont {Nori}}]{bliokh2014extraordinary}%
  \BibitemOpen
  \bibfield  {author} {\bibinfo {author} {\bibfnamefont {K.~Y.}\ \bibnamefont
  {Bliokh}}, \bibinfo {author} {\bibfnamefont {A.~Y.}\ \bibnamefont
  {Bekshaev}}, \ and\ \bibinfo {author} {\bibfnamefont {F.}~\bibnamefont
  {Nori}},\ }\href@noop {} {\bibfield  {journal} {\bibinfo  {journal} {Nat.
  Commun.}\ }\textbf {\bibinfo {volume} {5}},\ \bibinfo {pages} {3300}
  (\bibinfo {year} {2014}{\natexlab{a}})}\BibitemShut {NoStop}%
\bibitem [{\citenamefont {Bliokh}\ and\ \citenamefont
  {Nori}(2015)}]{Bliokh20151}%
  \BibitemOpen
  \bibfield  {author} {\bibinfo {author} {\bibfnamefont {K.~Y.}\ \bibnamefont
  {Bliokh}}\ and\ \bibinfo {author} {\bibfnamefont {F.}~\bibnamefont {Nori}},\
  }\href {\doibase http://dx.doi.org/10.1016/j.physrep.2015.06.003} {\bibfield
  {journal} {\bibinfo  {journal} {Phys. Rep.}\ }\textbf {\bibinfo {volume}
  {592}},\ \bibinfo {pages} {1 } (\bibinfo {year} {2015})}\BibitemShut
  {NoStop}%
\bibitem [{\citenamefont {Aiello}\ \emph {et~al.}(2015)\citenamefont {Aiello},
  \citenamefont {Banzer}, \citenamefont {Neugebauer},\ and\ \citenamefont
  {Leuchs}}]{Aiello2015}%
  \BibitemOpen
  \bibfield  {author} {\bibinfo {author} {\bibfnamefont {A.}~\bibnamefont
  {Aiello}}, \bibinfo {author} {\bibfnamefont {P.}~\bibnamefont {Banzer}},
  \bibinfo {author} {\bibfnamefont {M.}~\bibnamefont {Neugebauer}}, \ and\
  \bibinfo {author} {\bibfnamefont {G.}~\bibnamefont {Leuchs}},\ }\href@noop {}
  {\bibfield  {journal} {\bibinfo  {journal} {Nat. Photonics}\ }\textbf
  {\bibinfo {volume} {9}},\ \bibinfo {pages} {789} (\bibinfo {year}
  {2015})}\BibitemShut {NoStop}%
\bibitem [{\citenamefont {Rodriguez-Fortuno}\ \emph {et~al.}(2013)\citenamefont
  {Rodriguez-Fortuno}, \citenamefont {Marino}, \citenamefont {Ginzburg},
  \citenamefont {O’Connor}, \citenamefont {Martinez}, \citenamefont {Wurtz},\
  and\ \citenamefont {Zayats}}]{Zayats2013}%
  \BibitemOpen
  \bibfield  {author} {\bibinfo {author} {\bibfnamefont {F.}~\bibnamefont
  {Rodriguez-Fortuno}}, \bibinfo {author} {\bibfnamefont {G.}~\bibnamefont
  {Marino}}, \bibinfo {author} {\bibfnamefont {P.}~\bibnamefont {Ginzburg}},
  \bibinfo {author} {\bibfnamefont {D.}~\bibnamefont {O’Connor}}, \bibinfo
  {author} {\bibfnamefont {A.}~\bibnamefont {Martinez}}, \bibinfo {author}
  {\bibfnamefont {G.}~\bibnamefont {Wurtz}}, \ and\ \bibinfo {author}
  {\bibfnamefont {A.}~\bibnamefont {Zayats}},\ }\href@noop {} {\bibfield
  {journal} {\bibinfo  {journal} {Science}\ }\textbf {\bibinfo {volume}
  {340}},\ \bibinfo {pages} {328—} (\bibinfo {year} {2013})}\BibitemShut
  {NoStop}%
\bibitem [{\citenamefont {Petersen}\ \emph {et~al.}(2014)\citenamefont
  {Petersen}, \citenamefont {Volz},\ and\ \citenamefont
  {Rauschenbeutel}}]{Petersen2014}%
  \BibitemOpen
  \bibfield  {author} {\bibinfo {author} {\bibfnamefont {J.}~\bibnamefont
  {Petersen}}, \bibinfo {author} {\bibfnamefont {J.}~\bibnamefont {Volz}}, \
  and\ \bibinfo {author} {\bibfnamefont {A.}~\bibnamefont {Rauschenbeutel}},\
  }\href@noop {} {\bibfield  {journal} {\bibinfo  {journal} {Science}\ }\textbf
  {\bibinfo {volume} {346}},\ \bibinfo {pages} {67—} (\bibinfo {year}
  {2014})}\BibitemShut {NoStop}%
\bibitem [{\citenamefont {le~Feber}\ \emph {et~al.}(2015)\citenamefont
  {le~Feber}, \citenamefont {Rotenberg},\ and\ \citenamefont
  {Kuipers}}]{Kuipers2015}%
  \BibitemOpen
  \bibfield  {author} {\bibinfo {author} {\bibfnamefont {B.}~\bibnamefont
  {le~Feber}}, \bibinfo {author} {\bibfnamefont {N.}~\bibnamefont {Rotenberg}},
  \ and\ \bibinfo {author} {\bibfnamefont {L.}~\bibnamefont {Kuipers}},\
  }\href@noop {} {\bibfield  {journal} {\bibinfo  {journal} {Nat. Commun.}\
  }\textbf {\bibinfo {volume} {6}},\ \bibinfo {pages} {6695} (\bibinfo {year}
  {2015})}\BibitemShut {NoStop}%
\bibitem [{\citenamefont {S{\"o}llner}\ \emph {et~al.}(2015)\citenamefont
  {S{\"o}llner}, \citenamefont {Mahmoodian}, \citenamefont {Hansen},
  \citenamefont {Midolo}, \citenamefont {Javadi}, \citenamefont
  {Kir{\v{s}}ansk{\.e}}, \citenamefont {Pregnolato}, \citenamefont {El-Ella},
  \citenamefont {Lee}, \citenamefont {Song}, \citenamefont {Stobe},\ and\
  \citenamefont {Lodahl}}]{Sollner2015}%
  \BibitemOpen
  \bibfield  {author} {\bibinfo {author} {\bibfnamefont {I.}~\bibnamefont
  {S{\"o}llner}}, \bibinfo {author} {\bibfnamefont {S.}~\bibnamefont
  {Mahmoodian}}, \bibinfo {author} {\bibfnamefont {S.~L.}\ \bibnamefont
  {Hansen}}, \bibinfo {author} {\bibfnamefont {L.}~\bibnamefont {Midolo}},
  \bibinfo {author} {\bibfnamefont {A.}~\bibnamefont {Javadi}}, \bibinfo
  {author} {\bibfnamefont {G.}~\bibnamefont {Kir{\v{s}}ansk{\.e}}}, \bibinfo
  {author} {\bibfnamefont {T.}~\bibnamefont {Pregnolato}}, \bibinfo {author}
  {\bibfnamefont {H.}~\bibnamefont {El-Ella}}, \bibinfo {author} {\bibfnamefont
  {E.~H.}\ \bibnamefont {Lee}}, \bibinfo {author} {\bibfnamefont {J.~D.}\
  \bibnamefont {Song}}, \bibinfo {author} {\bibfnamefont {S.}~\bibnamefont
  {Stobe}}, \ and\ \bibinfo {author} {\bibfnamefont {P.}~\bibnamefont
  {Lodahl}},\ }\href@noop {} {\bibfield  {journal} {\bibinfo  {journal} {Nat.
  Nanotechnol.}\ }\textbf {\bibinfo {volume} {10}},\ \bibinfo {pages} {775}
  (\bibinfo {year} {2015})}\BibitemShut {NoStop}%
\bibitem [{\citenamefont {Bliokh}\ \emph
  {et~al.}(2015{\natexlab{a}})\citenamefont {Bliokh}, \citenamefont
  {Smirnova},\ and\ \citenamefont {Nori}}]{bliokh2015quantum}%
  \BibitemOpen
  \bibfield  {author} {\bibinfo {author} {\bibfnamefont {K.~Y.}\ \bibnamefont
  {Bliokh}}, \bibinfo {author} {\bibfnamefont {D.}~\bibnamefont {Smirnova}}, \
  and\ \bibinfo {author} {\bibfnamefont {F.}~\bibnamefont {Nori}},\ }\href
  {\doibase 10.1126/science.aaa9519} {\bibfield  {journal} {\bibinfo  {journal}
  {Science}\ }\textbf {\bibinfo {volume} {348}},\ \bibinfo {pages} {1448}
  (\bibinfo {year} {2015}{\natexlab{a}})}\BibitemShut {NoStop}%
\bibitem [{\citenamefont {Gloppe}\ \emph {et~al.}(2014)\citenamefont {Gloppe},
  \citenamefont {Verlot}, \citenamefont {Dupont-Ferrier}, \citenamefont
  {Siria}, \citenamefont {Poncharal}, \citenamefont {Bachelier}, \citenamefont
  {Vincent},\ and\ \citenamefont {Arcizet}}]{Gloppe2014}%
  \BibitemOpen
  \bibfield  {author} {\bibinfo {author} {\bibfnamefont {A.}~\bibnamefont
  {Gloppe}}, \bibinfo {author} {\bibfnamefont {P.}~\bibnamefont {Verlot}},
  \bibinfo {author} {\bibfnamefont {E.}~\bibnamefont {Dupont-Ferrier}},
  \bibinfo {author} {\bibfnamefont {A.}~\bibnamefont {Siria}}, \bibinfo
  {author} {\bibfnamefont {P.}~\bibnamefont {Poncharal}}, \bibinfo {author}
  {\bibfnamefont {G.}~\bibnamefont {Bachelier}}, \bibinfo {author}
  {\bibfnamefont {P.}~\bibnamefont {Vincent}}, \ and\ \bibinfo {author}
  {\bibfnamefont {O.}~\bibnamefont {Arcizet}},\ }\href@noop {} {\bibfield
  {journal} {\bibinfo  {journal} {Nat. Nanotechnol.}\ }\textbf {\bibinfo
  {volume} {9}},\ \bibinfo {pages} {920} (\bibinfo {year} {2014})}\BibitemShut
  {NoStop}%
\bibitem [{\citenamefont {Neukirch}\ \emph {et~al.}(2015)\citenamefont
  {Neukirch}, \citenamefont {von Haartman}, \citenamefont {Rosenholm},\ and\
  \citenamefont {Vamivakas}}]{Neukirch2015}%
  \BibitemOpen
  \bibfield  {author} {\bibinfo {author} {\bibfnamefont {L.~P.}\ \bibnamefont
  {Neukirch}}, \bibinfo {author} {\bibfnamefont {E.}~\bibnamefont {von
  Haartman}}, \bibinfo {author} {\bibfnamefont {J.~M.}\ \bibnamefont
  {Rosenholm}}, \ and\ \bibinfo {author} {\bibfnamefont {A.~N.}\ \bibnamefont
  {Vamivakas}},\ }\href@noop {} {\bibfield  {journal} {\bibinfo  {journal}
  {Nat. Photonics}\ }\textbf {\bibinfo {volume} {9}},\ \bibinfo {pages} {653}
  (\bibinfo {year} {2015})}\BibitemShut {NoStop}%
\bibitem [{\citenamefont {Belotelov}\ and\ \citenamefont
  {Zvezdin}(2012)}]{TMOKE1}%
  \BibitemOpen
  \bibfield  {author} {\bibinfo {author} {\bibfnamefont {V.}~\bibnamefont
  {Belotelov}}\ and\ \bibinfo {author} {\bibfnamefont {A.}~\bibnamefont
  {Zvezdin}},\ }\href@noop {} {\bibfield  {journal} {\bibinfo  {journal} {Phys.
  Rev. B}\ }\textbf {\bibinfo {volume} {86}},\ \bibinfo {pages} {155133}
  (\bibinfo {year} {2012})}\BibitemShut {NoStop}%
\bibitem [{\citenamefont {Bliokh}\ \emph
  {et~al.}(2015{\natexlab{b}})\citenamefont {Bliokh}, \citenamefont
  {Rodriguez-Fortuno}, \citenamefont {Nori},\ and\ \citenamefont
  {Zayats}}]{Bliokh2015SOI}%
  \BibitemOpen
  \bibfield  {author} {\bibinfo {author} {\bibfnamefont {K.}~\bibnamefont
  {Bliokh}}, \bibinfo {author} {\bibfnamefont {F.}~\bibnamefont
  {Rodriguez-Fortuno}}, \bibinfo {author} {\bibfnamefont {F.}~\bibnamefont
  {Nori}}, \ and\ \bibinfo {author} {\bibfnamefont {A.}~\bibnamefont
  {Zayats}},\ }\href@noop {} {\bibfield  {journal} {\bibinfo  {journal} {Nat.
  Photon.}\ }\textbf {\bibinfo {volume} {9}},\ \bibinfo {pages} {796—}
  (\bibinfo {year} {2015}{\natexlab{b}})}\BibitemShut {NoStop}%
\bibitem [{\citenamefont {Caldwell}\ \emph {et~al.}(2014)\citenamefont
  {Caldwell}, \citenamefont {Kretinin}, \citenamefont {Chen}, \citenamefont
  {Giannini}, \citenamefont {Fogler}, \citenamefont {Francescato},
  \citenamefont {Ellis}, \citenamefont {Tischler}, \citenamefont {Woods},
  \citenamefont {Giles}, \citenamefont {Hong}, \citenamefont {Watanabe},
  \citenamefont {Taniguchi}, \citenamefont {Maier},\ and\ \citenamefont
  {Novoselov}}]{caldwell2014sub}%
  \BibitemOpen
  \bibfield  {author} {\bibinfo {author} {\bibfnamefont {J.~D.}\ \bibnamefont
  {Caldwell}}, \bibinfo {author} {\bibfnamefont {A.~V.}\ \bibnamefont
  {Kretinin}}, \bibinfo {author} {\bibfnamefont {Y.}~\bibnamefont {Chen}},
  \bibinfo {author} {\bibfnamefont {V.}~\bibnamefont {Giannini}}, \bibinfo
  {author} {\bibfnamefont {M.~M.}\ \bibnamefont {Fogler}}, \bibinfo {author}
  {\bibfnamefont {Y.}~\bibnamefont {Francescato}}, \bibinfo {author}
  {\bibfnamefont {C.~T.}\ \bibnamefont {Ellis}}, \bibinfo {author}
  {\bibfnamefont {J.~G.}\ \bibnamefont {Tischler}}, \bibinfo {author}
  {\bibfnamefont {C.~R.}\ \bibnamefont {Woods}}, \bibinfo {author}
  {\bibfnamefont {A.~J.}\ \bibnamefont {Giles}}, \bibinfo {author}
  {\bibfnamefont {M.}~\bibnamefont {Hong}}, \bibinfo {author} {\bibfnamefont
  {K.}~\bibnamefont {Watanabe}}, \bibinfo {author} {\bibfnamefont
  {T.}~\bibnamefont {Taniguchi}}, \bibinfo {author} {\bibfnamefont {S.~A.}\
  \bibnamefont {Maier}}, \ and\ \bibinfo {author} {\bibfnamefont {K.~S.}\
  \bibnamefont {Novoselov}},\ }\href@noop {} {\bibfield  {journal} {\bibinfo
  {journal} {Nat. Commun.}\ }\textbf {\bibinfo {volume} {5}},\ \bibinfo {pages}
  {5221} (\bibinfo {year} {2014})}\BibitemShut {NoStop}%
\bibitem [{\citenamefont {Li}\ \emph {et~al.}(2015)\citenamefont {Li},
  \citenamefont {Lewin}, \citenamefont {Kretinin}, \citenamefont {Caldwell},
  \citenamefont {Novoselov}, \citenamefont {Taniguchi}, \citenamefont
  {Watanabe}, \citenamefont {Gaussmann},\ and\ \citenamefont
  {Taubner}}]{li2015hyperbolic}%
  \BibitemOpen
  \bibfield  {author} {\bibinfo {author} {\bibfnamefont {P.}~\bibnamefont
  {Li}}, \bibinfo {author} {\bibfnamefont {M.}~\bibnamefont {Lewin}}, \bibinfo
  {author} {\bibfnamefont {A.~V.}\ \bibnamefont {Kretinin}}, \bibinfo {author}
  {\bibfnamefont {J.~D.}\ \bibnamefont {Caldwell}}, \bibinfo {author}
  {\bibfnamefont {K.~S.}\ \bibnamefont {Novoselov}}, \bibinfo {author}
  {\bibfnamefont {T.}~\bibnamefont {Taniguchi}}, \bibinfo {author}
  {\bibfnamefont {K.}~\bibnamefont {Watanabe}}, \bibinfo {author}
  {\bibfnamefont {F.}~\bibnamefont {Gaussmann}}, \ and\ \bibinfo {author}
  {\bibfnamefont {T.}~\bibnamefont {Taubner}},\ }\href@noop {} {\bibfield
  {journal} {\bibinfo  {journal} {Nat. Commun.}\ }\textbf {\bibinfo {volume}
  {6}},\ \bibinfo {pages} {7507} (\bibinfo {year} {2015})}\BibitemShut
  {NoStop}%
\bibitem [{\citenamefont {Dai}\ \emph {et~al.}(2015)\citenamefont {Dai},
  \citenamefont {Ma}, \citenamefont {Andersen}, \citenamefont {Mcleod},
  \citenamefont {Fei}, \citenamefont {Liu}, \citenamefont {Wagner},
  \citenamefont {Watanabe}, \citenamefont {Taniguchi}, \citenamefont
  {Thiemens}, \citenamefont {Keilmann}, \citenamefont {Jarillo-Herrero},
  \citenamefont {Fogler},\ and\ \citenamefont
  {Basov}}]{dai2015subdiffractional}%
  \BibitemOpen
  \bibfield  {author} {\bibinfo {author} {\bibfnamefont {S.}~\bibnamefont
  {Dai}}, \bibinfo {author} {\bibfnamefont {Q.}~\bibnamefont {Ma}}, \bibinfo
  {author} {\bibfnamefont {T.}~\bibnamefont {Andersen}}, \bibinfo {author}
  {\bibfnamefont {A.}~\bibnamefont {Mcleod}}, \bibinfo {author} {\bibfnamefont
  {Z.}~\bibnamefont {Fei}}, \bibinfo {author} {\bibfnamefont {M.}~\bibnamefont
  {Liu}}, \bibinfo {author} {\bibfnamefont {M.}~\bibnamefont {Wagner}},
  \bibinfo {author} {\bibfnamefont {K.}~\bibnamefont {Watanabe}}, \bibinfo
  {author} {\bibfnamefont {T.}~\bibnamefont {Taniguchi}}, \bibinfo {author}
  {\bibfnamefont {M.}~\bibnamefont {Thiemens}}, \bibinfo {author}
  {\bibfnamefont {F.}~\bibnamefont {Keilmann}}, \bibinfo {author}
  {\bibfnamefont {P.}~\bibnamefont {Jarillo-Herrero}}, \bibinfo {author}
  {\bibfnamefont {M.~M.}\ \bibnamefont {Fogler}}, \ and\ \bibinfo {author}
  {\bibfnamefont {D.~N.}\ \bibnamefont {Basov}},\ }\href@noop {} {\bibfield
  {journal} {\bibinfo  {journal} {Nat. Commun.}\ }\textbf {\bibinfo {volume}
  {6}},\ \bibinfo {pages} {6963} (\bibinfo {year} {2015})}\BibitemShut
  {NoStop}%
\bibitem [{\citenamefont {Liu}\ \emph {et~al.}(2007)\citenamefont {Liu},
  \citenamefont {Lee}, \citenamefont {Xiong}, \citenamefont {Sun},\ and\
  \citenamefont {Zhang}}]{liu2007far}%
  \BibitemOpen
  \bibfield  {author} {\bibinfo {author} {\bibfnamefont {Z.}~\bibnamefont
  {Liu}}, \bibinfo {author} {\bibfnamefont {H.}~\bibnamefont {Lee}}, \bibinfo
  {author} {\bibfnamefont {Y.}~\bibnamefont {Xiong}}, \bibinfo {author}
  {\bibfnamefont {C.}~\bibnamefont {Sun}}, \ and\ \bibinfo {author}
  {\bibfnamefont {X.}~\bibnamefont {Zhang}},\ }\href@noop {} {\bibfield
  {journal} {\bibinfo  {journal} {Science}\ }\textbf {\bibinfo {volume}
  {315}},\ \bibinfo {pages} {1686} (\bibinfo {year} {2007})}\BibitemShut
  {NoStop}%
\bibitem [{\citenamefont {Ishii}\ \emph {et~al.}(2013)\citenamefont {Ishii},
  \citenamefont {Kildishev}, \citenamefont {Narimanov}, \citenamefont
  {Shalaev},\ and\ \citenamefont {Drachev}}]{ishii2013sub}%
  \BibitemOpen
  \bibfield  {author} {\bibinfo {author} {\bibfnamefont {S.}~\bibnamefont
  {Ishii}}, \bibinfo {author} {\bibfnamefont {A.~V.}\ \bibnamefont
  {Kildishev}}, \bibinfo {author} {\bibfnamefont {E.}~\bibnamefont
  {Narimanov}}, \bibinfo {author} {\bibfnamefont {V.~M.}\ \bibnamefont
  {Shalaev}}, \ and\ \bibinfo {author} {\bibfnamefont {V.~P.}\ \bibnamefont
  {Drachev}},\ }\href@noop {} {\bibfield  {journal} {\bibinfo  {journal} {Laser
  Photonics Rev.}\ }\textbf {\bibinfo {volume} {7}},\ \bibinfo {pages} {265}
  (\bibinfo {year} {2013})}\BibitemShut {NoStop}%
\bibitem [{\citenamefont {Iorsh}\ \emph {et~al.}(2013)\citenamefont {Iorsh},
  \citenamefont {Mukhin}, \citenamefont {Shadrivov}, \citenamefont {Belov},\
  and\ \citenamefont {Kivshar}}]{iorsh2013hyperbolic}%
  \BibitemOpen
  \bibfield  {author} {\bibinfo {author} {\bibfnamefont {I.~V.}\ \bibnamefont
  {Iorsh}}, \bibinfo {author} {\bibfnamefont {I.~S.}\ \bibnamefont {Mukhin}},
  \bibinfo {author} {\bibfnamefont {I.~V.}\ \bibnamefont {Shadrivov}}, \bibinfo
  {author} {\bibfnamefont {P.~A.}\ \bibnamefont {Belov}}, \ and\ \bibinfo
  {author} {\bibfnamefont {Y.~S.}\ \bibnamefont {Kivshar}},\ }\href@noop {}
  {\bibfield  {journal} {\bibinfo  {journal} {Phys. Rev. B}\ }\textbf {\bibinfo
  {volume} {87}},\ \bibinfo {pages} {075416} (\bibinfo {year}
  {2013})}\BibitemShut {NoStop}%
\bibitem [{\citenamefont {Chshelokova}\ \emph {et~al.}(2012)\citenamefont
  {Chshelokova}, \citenamefont {Kapitanova}, \citenamefont {Poddubny},
  \citenamefont {Filonov}, \citenamefont {Slobozhanyuk}, \citenamefont
  {Kivshar},\ and\ \citenamefont {Belov}}]{chshelokova2012hyperbolic}%
  \BibitemOpen
  \bibfield  {author} {\bibinfo {author} {\bibfnamefont {A.~V.}\ \bibnamefont
  {Chshelokova}}, \bibinfo {author} {\bibfnamefont {P.~V.}\ \bibnamefont
  {Kapitanova}}, \bibinfo {author} {\bibfnamefont {A.~N.}\ \bibnamefont
  {Poddubny}}, \bibinfo {author} {\bibfnamefont {D.~S.}\ \bibnamefont
  {Filonov}}, \bibinfo {author} {\bibfnamefont {A.~P.}\ \bibnamefont
  {Slobozhanyuk}}, \bibinfo {author} {\bibfnamefont {Y.~S.}\ \bibnamefont
  {Kivshar}}, \ and\ \bibinfo {author} {\bibfnamefont {P.~A.}\ \bibnamefont
  {Belov}},\ }\href@noop {} {\bibfield  {journal} {\bibinfo  {journal} {J.
  Appl. Phys.}\ }\textbf {\bibinfo {volume} {112}},\ \bibinfo {pages} {073116}
  (\bibinfo {year} {2012})}\BibitemShut {NoStop}%
\bibitem [{\citenamefont {Bliokh}\ \emph
  {et~al.}(2014{\natexlab{b}})\citenamefont {Bliokh}, \citenamefont {Kivshar},\
  and\ \citenamefont {Nori}}]{Bliokh2014}%
  \BibitemOpen
  \bibfield  {author} {\bibinfo {author} {\bibfnamefont {K.~Y.}\ \bibnamefont
  {Bliokh}}, \bibinfo {author} {\bibfnamefont {Y.~S.}\ \bibnamefont {Kivshar}},
  \ and\ \bibinfo {author} {\bibfnamefont {F.}~\bibnamefont {Nori}},\
  }\href@noop {} {\bibfield  {journal} {\bibinfo  {journal} {Phys. Rev. Lett.}\
  }\textbf {\bibinfo {volume} {113}},\ \bibinfo {pages} {033601} (\bibinfo
  {year} {2014}{\natexlab{b}})}\BibitemShut {NoStop}%
\bibitem [{\citenamefont {Mikhailov}\ and\ \citenamefont
  {Ziegler}(2007)}]{Mikhailov2007}%
  \BibitemOpen
  \bibfield  {author} {\bibinfo {author} {\bibfnamefont {S.~A.}\ \bibnamefont
  {Mikhailov}}\ and\ \bibinfo {author} {\bibfnamefont {K.}~\bibnamefont
  {Ziegler}},\ }\href {\doibase 10.1103/PhysRevLett.99.016803} {\bibfield
  {journal} {\bibinfo  {journal} {Phys. Rev. Lett.}\ }\textbf {\bibinfo
  {volume} {99}},\ \bibinfo {pages} {016803} (\bibinfo {year}
  {2007})}\BibitemShut {NoStop}%
\end{thebibliography}%
	
\end{document}